\documentclass[aps,twocolumn,showpacs,preprintnumbers,amsmath,amssymb]{revtex4}
\usepackage{graphicx}
\begin{document}
\title{Phase transition in ultrathin magnetic films with long range interactions}
\author{M. Rapini} \email{mrapini@fisica.ufmg.br}
\affiliation{Laborat\'orio de Simula\c{c}\~ao - Departamento de
F\'{\i}sica - ICEX - UFMG 30123-970 Belo Horizonte - MG, Brazil}
\author{R.A. Dias}\email{radias@fisica.ufmg.br}
\affiliation{Laborat\'orio de Simula\c{c}\~ao - Departamento de
F\'{\i}sica - ICEX - UFMG 30123-970 Belo Horizonte - MG, Brazil}
\author{B.V. Costa} \email{bvc@fisica.ufmg.br}
\affiliation{Laborat\'orio de Simula\c{c}\~ao - Departamento de
F\'{\i}sica - ICEX - UFMG 30123-970 Belo Horizonte - MG, Brazil}
\begin{abstract}
Ultrathin magnetic films can be modeled as an anisotropic
Heisenberg model with long range dipolar interactions. It is
believed that the phase diagram presents three phases: A ordered
ferromagnetic phase ($I$), a phase characterized by a change from
out-of-plane to in-plane in the magnetization ($II$), and a high
temperature paramagnetic phase ($III$). It is claimed that the
border lines from phase $I$ to $III$ and $II$ to $III$ are of
second order and from $I$ to $II$ is first order. In the present
work we have performed a very careful Monte Carlo simulation of
the model. Our results strongly support that the line separating
phase $II$ and $III$ is of the $BKT$ type.
\end{abstract}
\maketitle

%
%
\section{Introduction}
Since the late 80's there has being an increasing interest in
ultrathin magnetic films
\cite{salamon,majkrzak,dutcher,gruenberg,saurenbach,allenspach}.
This interest is mainly associated to the development of
magnetic-non-magnetic multilayers for the purpose of giant
magnetoresistence applications \cite{levy}. In addition, experiments
on epitaxial magnetic layers have shown that a huge variety of
complex structures can develop in the system
\cite{chapman,daykin,johnston,hehn}. Rich magnetic domain structures
like stripes, chevrons, labyrinths and bubbles associated to the
competition between dipolar long range interactions and strong
anisotropies perpendicular to the plane of the film were observed
experimentally. A lot of theoretical work has been done on the
morphology and stability of these magnetic structures
\cite{chui,vedmedenko1,vedmedenko2}. Beside that, it has been
observed the existence of a switching transition from perpendicular
to in-plane ordering at low but finite temperature
\cite{pappas,allenspach2,rapini-costa-landau,santamaria}: at low
temperature the film magnetization is perpendicular to the film
surface, as temperature rises the magnetization flips to an in-plane
configuration. Eventually the out-of-plane and the in plane
magnetization become zero \cite{chui2}.

\begin{figure} [htbp]
\vspace{0.5cm}
\includegraphics[height=6.0cm,width=6.0cm]{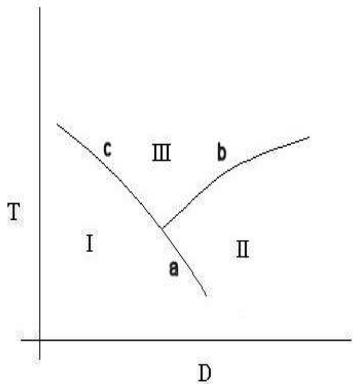}
\caption{A sketch of the phase diagram for the model, (eq. \ref{hamiltonian}).
Phase $I$ correspond to an
out-of-plane magnetization, phase $II$ has in-plane magnetization
and phase $III$ is paramagnetic. The border line between phase $I$
to phase $II$ is believed to be of first order and from region $I$
and $II$ to $III$ are both second order.}
\label{fig1}
\end{figure}

The general Hamiltonian describing a prototype for a ultrathin
magnetic film assumed to lay in the $xy$ plane is
\cite{santamaria}
\begin {eqnarray}\label{hamiltonian}
H = -J\sum_{<ij>} \vec{S_i}\cdot\vec{S_j} - A\sum_{i} {S_i^z}^2 +
\\ \nonumber
D\sum_{<ij>} \left[  \frac{\vec{S_i}\cdot\vec{S_j}}{{r}_{ij}^3} -
3\frac{\left( \vec{S_{i}} \cdot \vec{r}_{ij} \right) \cdot \left(
\vec{S_{j}} \cdot \vec{r}_{ij} \right)} {{r}_{ij}^5} \right] .
\end{eqnarray}
Here $J$ is an exchange interaction which is assumed to be nonzero
only for nearest-neighbor interaction, $D$ is the dipolar coupling
parameter, $A$ is a single-ion anisotropy and
$\vec{r}_{ij}=\vec{r}_j - \vec{r}_i$ where $\vec{r}_i$ stands for
lattice vectors. The structures developed in the system depend on
the sample geometry and size. Several situations are discussed in
reference
\cite{vedmedenko2} and citations therein. \\
Although the structures developed in the system are well known the
phase diagram of the model is still being studied. There are several
possibilities since we can combine the parameters in many ways. We
want to analyze the case $J > 0$ in some interesting situations. A
more detailed analysis covering the entire space of parameters is
under consideration.
\begin{itemize}
\item Case $D=0$ \\
For $D=0$ we recover the two dimensional ($2d$) anisotropic
Heisenberg model. The isotropic case, $ A = 0$, is known to
present no transition \cite{mermin-wagner}. For $A > 0$ the model
is in the Ising universality class \cite{Ising} undergoing a
order-disorder phase transition. If $A < 0$ the model is in the
$XY$ universality class. In this case it is known to have a
Berezinskii-Kosterlitz-Thouless ($BKT$) phase transition which is
characterized by a vortex-anti-vortex unbinding, with no true long
range order \cite{berezinskii,kosterliz-thouless,teitel,kogut}.
\item Case $D \neq 0$ \\
In this case, there is a competition between the dipolar and the
anisotropic terms. If $D$ is small compared to $A$ we can expect
the system to have an Ising behavior. If $D$ is not too small we
can expect a transition of the spins from out-of-plane to in-plane
configuration \cite{santamaria}. For large enough $D$ out-of-plane
configurations become instable such that, the system lowers its
energy by turning the spins into an in-plane anti-ferromagnetic
arrangement.
\end{itemize}
Earlier works on this model which discuss the phase diagram were
mostly done using renormalization group approach and numerical
Monte Carlo simulation \cite{santamaria,chui2,sak}. They agree
between themselves in the main features. The phase diagram for
fixed $A$ and $J=1$ is schematically shown in figure \ref{fig1}
in the space ($D,T$).
From Monte Carlo (MC) results it is found that there are three
regions labelled in the figure \ref{fig1} as $I, II$ and $III$.
Phase $I$ correspond to an out-of-plane magnetization, phase $II$
has in-plane magnetization and phase $III$ is paramagnetic. The
border line between phase $I$ to phase $II$ is believed to be of
first order and from region $I$ and $II$ to $III$ are both second
order.

Although, the different results agree between themselves about the
character of the different regions, much care has to be taken
because they were obtained by using a cut-off, $r_c$, in the dipolar
term. The long range character of the potential is lost,
consequently we can expect a line of $BKT$ transition coming from
region $II$ to region $III$. It is characterized by having no true
long range order. This lack of long range order is prevented by the
Mermin-Wagner theorem \cite{mermin-wagner}. The $BKT$ phase
transition is  an unusual magnetic phase transition characterized by
the unbinding of pairs of topological excitations named
vortex-anti-vortex
\cite{berezinskii,kosterliz-thouless,teitel,kogut,sak,evaristo1,evaristo3}.
A vortex (Anti-vortex) is a topological excitation in which spins on
a closed path around the excitation core precess by $2 \pi$ ($-2
\pi$). Above $T_{BKT}$ the correlation length behaves as $\xi
\approx \exp(bt^{-1/2})$, with $t\equiv (T-T_{BKT})/T_{BKT}$ and
$\xi \rightarrow \infty$ below $T_{BKT}$.

In this work we use MC simulations to investigate the model
defined by equation \ref{hamiltonian}. We use a cutoff, $r_c$, in
the dipolar interaction. Our results strongly suggest that the
transition between regions $II$ and $III$ is in the $BKT$
universality class, instead of second order, as found in earlier
works.

\section{Simulation Background}

Our simulations are done in a square lattice of volume $L \times
L$ ($L=10,20,30,40,50,80$) with periodic boundary conditions. We
use the Monte-Carlo method with the Metropolis algorithm
\cite{evaristo1,mc1,mc2,landau-binder}. To treat the dipole term
we use a cut-off $r_c=5a$, where $a$ is the lattice spacing, as
suggested in the work of Santamaria and co-workers
\cite{santamaria}.

We have performed the simulations for temperatures in the range
$0.3\leq T \leq 1.2$ at intervals of $\Delta T = 0.1$. When
necessary this temperature interval is reduced to $\Delta T =
0.01$. For every temperature the first $5 \times 10^5$ MC steps
per spin were used to lead the system to equilibrium. The next
$10^6$ configurations were used to calculate thermal averages of
thermodynamical quantities of interest. We have divided these last
$10^6$ configurations in $20$ bins from which the error bars are
estimated from the standard deviation of the averages over these
twenty runs. The single-site anisotropy constant was fixed as A =
1.0 while the $D$ parameter was set to $0.10, 0.15$ and $0.20$. In
this work the energy is measured in units of $JS^2$ and
temperature in units of $JS^2/k_B$, where $k_B$ is the Boltzmann
constant.

To estimate the transition temperatures we use finite-size-scaling
(FSS) analysis to the results of our MC simulations. In the
following we summarize the main FSS properties. If the reduced
temperature is $t = (T-T_c)/T$, the singular part of the free
energy, is given by
\begin{equation}\label{free-energy}
    F(L,T)= L^{-(2- \alpha)/\nu}{\cal F}(tL^{1/\nu})
\end{equation}
Appropriate differentiation of $F$ yields the various
thermodynamic properties. For an order disorder transition exactly
at $T_c$ the magnetization, $M$, susceptibility, $\xi$ and
specific heat, $C$, behave respectively as
\cite{landau-binder,privman}.
\begin{eqnarray}\label{scaling}
  M & \propto L^{-\beta/\nu} \nonumber \\
  \chi & \propto L^{-\gamma/\nu}  \\
  C & \propto L^{-\alpha/\nu}. \nonumber
\end{eqnarray}
In addition to these an important quantity is the forth order
Binder's cumulant
\begin{equation}\label{cumulant}
    U_4 = 1 - \frac{<m^4>}{3<m^2>^2}.
\end{equation}
For large enough $L$, curves for $U_4(T)$ cross the same point $U^*$
at $T=T_c$.
For a $BKT$ transition the quantities defined above behave in a
different way. Due to the Mermin-Wagner theorem there is no
spontaneous  magnetization for any finite temperature. The
specific heat present a peak at a temperature which is slightly
higher than $T_{BKT}$. Beside that, the peak height does not
depends on $L$. Because models presenting  a $BKT$ transition have
an entire critical region, the curves for $U_4(L)$ just coincide
inside that region presenting no crosses at all. Below we present
MC results for three typical regions. When not indicated the error
bars are smaller than the symbol sizes.
\section{Simulation Results}
\subsection{Case D=0.1}
\begin{figure} [htbp]
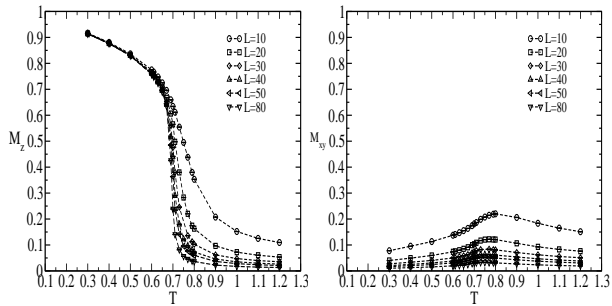

\vspace{0.5cm}
\includegraphics[height=4.0cm,width=4.0cm]{fig2.eps}\includegraphics[height=4.0cm,width=4.0cm]{fig3.eps}
\caption{Out-of-plane (left) and in-plane (right) magnetization
for $D=0.1$. The ground state is ferromagnetic. There is no
in-plane spontaneous magnetization.} \label{md=01}
\end{figure}

\begin{figure} [htbp]
\vspace{0.5cm}
\includegraphics[height=6.0cm,width=6.0cm]{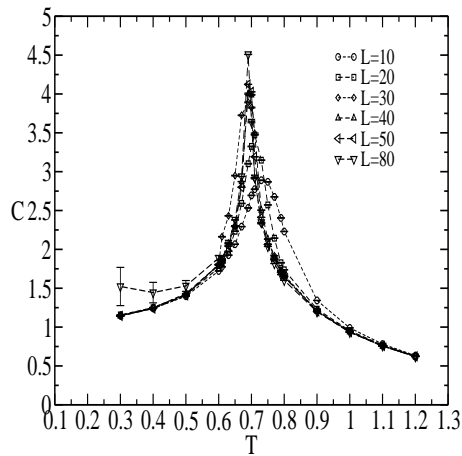}
\caption{Specific heat as function of temperature for $D=0.1$.}
\label{c-d=01}
\end{figure}

\begin{figure} [htbp]
\vspace{0.5cm}
\includegraphics[height=6.0cm,width=6.0cm]{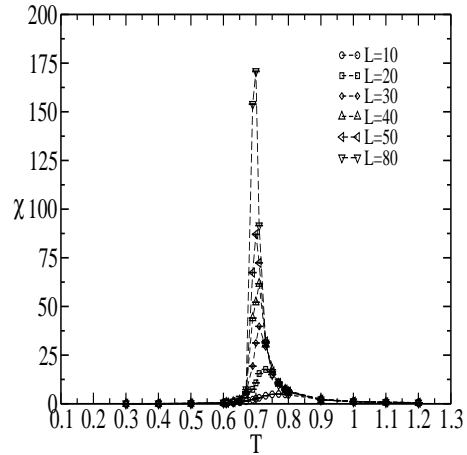}
\caption{Out-of-plane susceptibility as function of temperature for
$D=0.1$.} \label{xi-d=01}
\end{figure}

\begin{figure} [htbp]
\vspace{0.5cm}
\includegraphics[height=6.0cm,width=6.0cm]{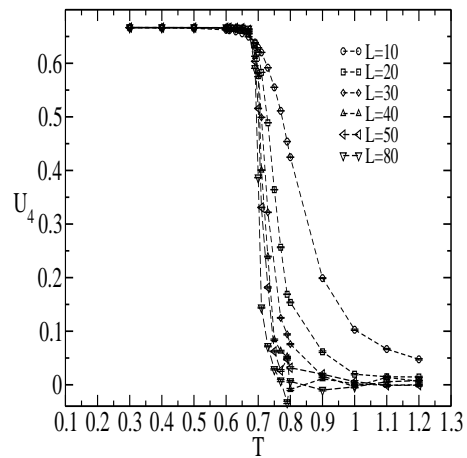}
\caption{Binder's cumulant as function of temperature for
$D=0.1$.} \label{u4-d=01}
\end{figure}

For $D=0.1$ we measured the dependence of the out-of-plane
magnetization, $M_z$, and the in-plane magnetization , $M_{xy}$, as
a function of temperature for several values of $L$ (See figure
\ref{md=01}).
The figures indicate that in the ground state the system is
aligned in the $z$ direction. Approximately at $T \approx 0.70$
the $M_z$ magnetization goes to zero, which gives a rough estimate
of the critical temperature. The in-plane magnetization has a
small peak close to $T \approx 0.70$. However, the height of the
peak diminishes as $L$ grows, in a clear indicative that it is a
finite size artifice. The behavior of the specific heat,
susceptibility and Binder's cumulant, are shown in figures
\ref{c-d=01}, \ref{xi-d=01} and \ref{u4-d=01} respectively. The
results indicate a order-disorder phase transition in clear
agreement with references \cite{pappas,allenspach2,
santamaria,chui2}.
\begin{figure} [htbp]
\vspace{0.5cm}
\includegraphics[height=6.0cm,width=6.0cm]{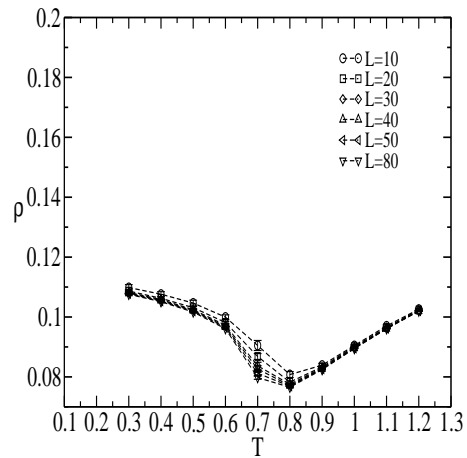}
\caption{Vortex density in the $xy$ plane for $D=0.1$.}
\label{rho-d=01}
\end{figure}
The vortex density in the $xy$ plane (Figure \ref{rho-d=01}) has a
very shallow minimum near the estimated critical temperature which
is almost independent of the lattice size. The growth of the
number of vortices when the temperature is decreased is related to
the disordering in the plane when the magnetic moments tend to be
in the $z$ direction. We have performed a finite size scaling
analysis of the data above by plotting the temperature $T_c^L$ as
a function of $1/L$ for the specific heat, the susceptibility and
the crosses of the forth-order cumulant. The results are shown in
the table \ref{table1}. By linear regression we have obtained the
critical temperature as $T_c^{\infty}= 0.682(2)$. An analysis of
the behavior of the maxima of the specific heat, $C_{max}$, (See
figure \ref{fssheat}) as a function of the lattice size shows that
it behaves as $C_{max} \propto \ln L$, indicating a second order
phase transition. In the phase diagram we crossed the second order
line labelled $c$.
\begin{table}[htbp]
\begin{tabular}{ccccccccc}
\\\hline \hline L & 10 & 20 & 30 & 40 & 50 & 80&\\\hline
$C$ & 0.735  & 0.711 & 0.695  & 0.693  & 0.690
 & 0.689 & \\
$\chi$& 0.771& 0.729& 0.710&0.707&0.700& 0.697 &\\
$U_4$& 0.675& 0.673 & 0.673 & 0.673 & 0.673 & - &\\ \hline
\end{tabular}
\caption{Critical temperature $T_c^L$ of the specific heat, $C$,
susceptibility, $\chi$, and the crosses of the fourth-order Binder's
cumulante $U_4$ as a function of the lattice size $L$. Data are for
$D=0.10$}
 \label{table1}
\end{table}
\subsection{Case D=0.15}
\begin{figure} [htbp]
\vspace{0.5cm}
\includegraphics[height=6.0cm,width=6.0cm]{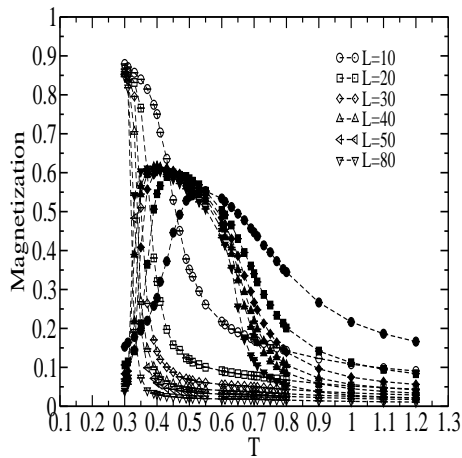}
\caption{$M_z$ and $M_{xy}$ (open and full symbols respectively)
for $D=0.15$.} \label{m-d=015}
\end{figure}

\begin{figure} [htbp]
\vspace{0.5cm}
\includegraphics[height=6.0cm,width=6.0cm]{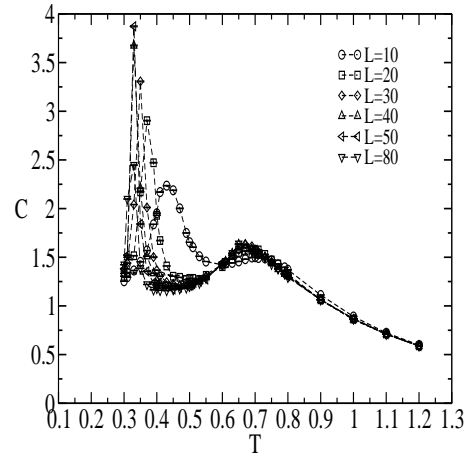}
\caption{Specific heat for $D=0.15$.}
 \label{c-d=015}
\end{figure}

In this region of the parameters, it was observed a transition from
an out-of-plane ordering at low temperatures to an in-plane
configuration as described by the magnetization behavior shown in
Fig. \ref{m-d=015}. We show $M_z$ and $M_{xy}$ in the same figure
for comparison. The out-of-plane magnetization goes to zero at $T
\approx 0.35$ while an in-plane magnetization sets in. This
phenomenon has already been reported experimentally \cite{pappas,
allenspach2} and it is due to the competition between the easy axis
anisotropy and the dipolar interaction.
The specific heat curve presents two peaks (See figure
\ref{c-d=015}.). The peak at low temperature is pronounced and is
centered in the temperature in which occurs the rapid decrease of
the in-plane magnetization, $T_1 \approx 0.35$. The second peak
appears at $T_2 \approx 0.65$ and seems to be independent of the
lattice size.
\begin{figure} [htbp]
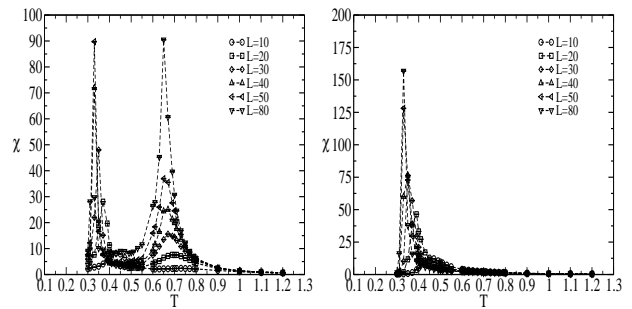

\vspace{0.5cm}
\includegraphics[height=4.0cm,width=4.0cm]{fig10.eps}
\includegraphics[height=4.0cm,width=4.0cm]{fig11.eps}
\caption{In-plane (left) and out-of-plane (right) susceptibility
for $D=0.15$.}
 \label{xi-d=015}
\end{figure}
In the figure \ref{xi-d=015} we show the in-plane and out-of-plane
susceptibilities. The out-of-plane susceptibility presents a single
peak close to $T_1 \approx 0.35$. The in-plane susceptibility has a
maxima at $T_2 \approx 0.65$ beside the peak at $T_1$, indicating
two phase transitions. The Binder's cumulant for the in-plane and
out-of-plane magnetization are shown in figures \ref{uxy4-d=015}.
Except for the case $L=10$ the curves for different values of the
lattice size do not cross each other indicating a $BKT$ transition
at $T \approx T_2$. Beside that, the in-plane cumulant has a minimum
at $T \approx T_1$, which is a characteristic of a first order phase
transition \cite{landau-binder,privman}.
\begin{figure} [htbp]
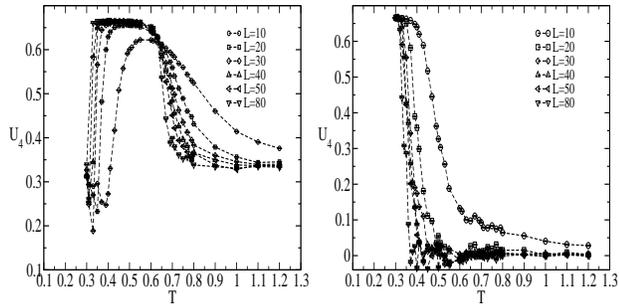

\vspace{0.5cm}
\includegraphics[height=4.0cm,width=4.0cm]{fig12.eps}
\includegraphics[height=4.0cm,width=4.0cm]{fig13.eps}
\caption{In-plane (left) and out-of-plane (right) Binder's
cumulant as function of temperature for $D=0.15$. Observe that the
in-plane cumulant has a minimum at $T \approx 0.35$ indicating a
first order phase transition. After the minimum the curves do not
cross each other having the same behavior (Except the spurious
case $L=10$.) up to $T \approx 0.65$ when they go apart. That is
an indication of a $BKT$ phase transition.} \label{uxy4-d=015}
\end{figure}

\begin{figure} [htbp]
\vspace{0.5cm}
\includegraphics[height=6.0cm,width=6.0cm]{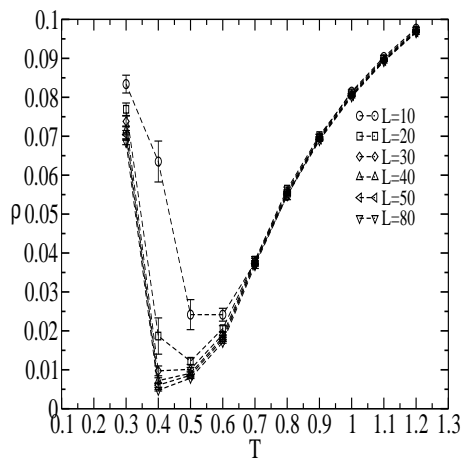}
\caption{Vortex density as function of temperature for $D=0.15$.}
\label{rho-d=015}
\end{figure}

The vortex density is shown in figure \ref{rho-d=015}. Its
behavior is similar to that one shown in figure \ref{rho-d=01}.
In the phase diagram we crossed the region I to II ($T_1$) and II
to III ($T_2$). The maxima of the specific heat are shown in
figure \ref{fssheat} as a function of $L$. It is clear that after
a transient behavior it remains constant indicating a $BKT$
transition. A $FSS$ analysis of the susceptibility (see table
\ref{table2} ) gives the $BKT$ temperature
$T_{BKT}^{\infty}=0.613(5)$.
\begin{table}[htbp]
\begin{tabular}{ccccccccc}
\\\hline \hline $T_c^L$ & 0.729  & 0.698 & 0.678& 0.670  & 0.650  &
0.638&  \\
L & 10 & 20 & 30 & 40 & 50 & 80&\\
\hline\\
\end{tabular}
\caption{Critical temperature $T_c^L$ as a function of the linear
size $L$ for the susceptibility $\chi$.} \label{table2}
\end{table}
In the phase diagram we crossed the first order line labelled $a$
($T_1$) and the line labelled $b$ ($T_2$).
\subsection{Case D=0.20}
\begin{figure} [htbp]
\vspace{0.5cm}
\includegraphics[height=6.0cm,width=6.0cm]{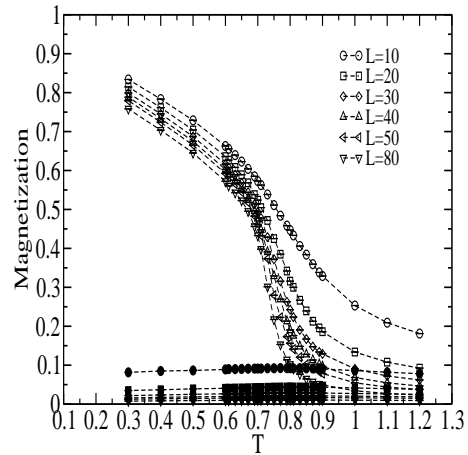}
\caption{$M_{xy}$ and $M_{z}$ (open and full symbols respectively)
for $D=0.2$.} \label{mxy-d=02}
\end{figure}

\begin{figure} [htbp]
\vspace{0.5cm}
\includegraphics[height=6.0cm,width=6.0cm]{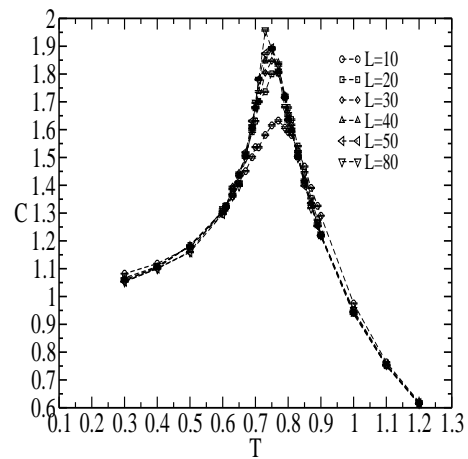}
\caption{Specific heat for $D=0.2$.The line is a guide to the
eyes.} \label{c-d=02}
\end{figure}

In figure \ref{mxy-d=02} we show the in-plane and out-of-plane
magnetization curves for several lattice sizes and $D=0.20$. We
observe that as the lattice size $L$ goes from $L=10$ to $L=80$,
both magnetization decrease. It can be inferred that as the system
approaches the thermodynamic limit, the net magnetization should
be zero. Therefore, the system does not present finite
magnetization for any temperature $T \neq 0$.
The specific heat (Figure \ref{c-d=02}) presents a maximum at $T
\approx 0.75$. The curves are for different values of $L$. We
observe that the position of the maxima and their heights are not
affected by the lattice size, all points falling in the same curve.
\begin{figure} [htbp]
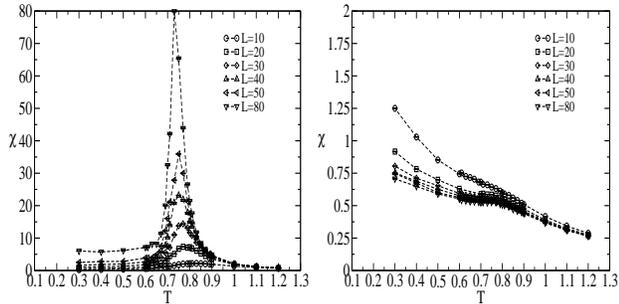

\vspace{0.5cm}
\includegraphics[height=4.0cm,width=4.0cm]{fig17.eps}
\includegraphics[height=4.0cm,width=4.0cm]{fig18.eps}
\caption{In-plane and out-of-plane susceptibility for $D=0.2$.}
\label{xi_z-d=02}
\end{figure}
\begin{figure} [htbp]
\vspace{0.5cm}
\includegraphics[height=6.0cm,width=6.0cm]{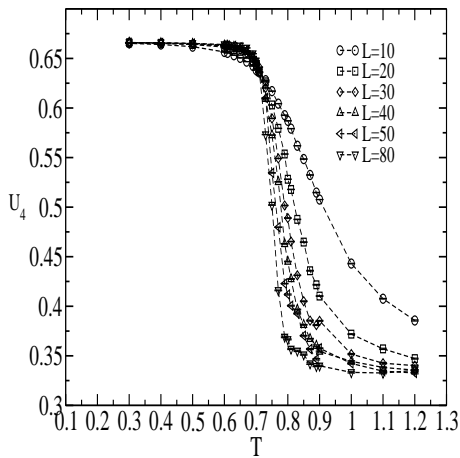}
\caption{Fourth-order in-plane cumulant for $D=0.2$.}
\label{uxy4-d=02}
\end{figure}
\begin{figure} [htbp]
\vspace{0.5cm}
\includegraphics[height=6.0cm,width=6.0cm]{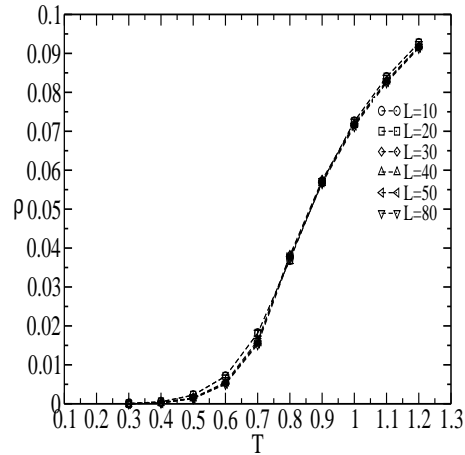}
\caption{Vortex density in the $xy$ plane for $D=0.2$.}
\label{rho-d=02}
\end{figure}
\begin{table}
\begin{tabular}{ccccccccc}
\\\hline \hline $T_c^L$ & 0.829  & 0.781 & 0.768& 0.753  & 0.750  &
0.729&  \\
L & 10 & 20 & 30 & 40 & 50 & 80&\\
\hline\\
\end{tabular}
\caption{Critical temperature $T_c^L$ as a function of the linear
size $L$ for the susceptibility $\chi$.}
 \label{table3}
\end{table}
\begin{figure} [htbp]
\vspace{0.5cm}
\includegraphics[height=6.0cm,width=6.0cm]{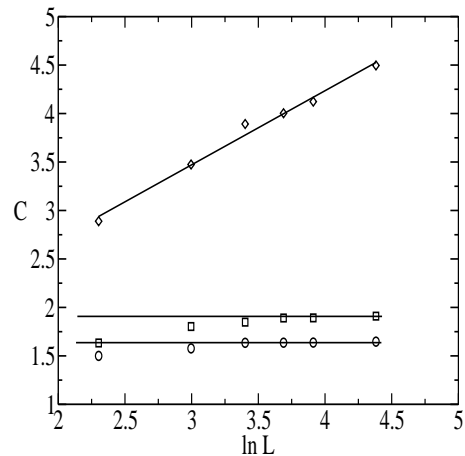}
\caption{Maxima of the specific heat as a function of the lattice
size.} \label{fssheat}
\end{figure}

In the figure \ref{xi_z-d=02} we show the in-plane and out-of-plane
susceptibilities respectively. $\chi^{zz}$ does not present any
critical behavior. $\chi^{xy}$ presents a peak which increases with
$L$. For the Binder's cumulant there is no unique cross of the
curves. (Except for the $L=10$ curve, which is considered too small
to be taken in to account.). This behavior indicates a $BKT$
transition at $T_{BKT} \approx 0.63$. The vortex density, shown in
figure \ref{uxy4-d=02} is almost independent on the lattice size. In
addition, we did a $FSS$ analysis of the susceptibility (see table
\ref{table3}) and the maxima of the specific heat. The specific heat
is shown in figure \ref{fssheat}. Its behavior indicates a $BKT$
transition. The analysis of the susceptibility gives
$T_{BKT}^{\infty}=0.709(5)$. In the phase diagram we crossed the
line labelled $b$. In our results we could not detect any other
transition for $D=0.20$, indicating that: The line labelled $a$ ends
somewhere in between $ 0.15 < D < 0.20$ or the crossing at $a$
occurs at a lower temperature ($T < 0.30$) outside the range of our
simulated data.
\section{Conclusions}
In earlier studies several authors have claimed that the model for
ultrathin magnetic films defined by the equation \ref{hamiltonian}
presents three phases. Referring to figure \ref{fig1} it is believed
that the line labelled $a$ is of first order. The line $b$ and $c$
are of second order. Those results were obtained by introducing a
cut off in the long range interaction of the hamiltonian. In the
present work we have used a numerical Monte Carlo approach to study
the phase diagram of the model for $J=A=1$ and $D = 0.10, 0.15$ and
$0.20$. In order to compare our results to those discussed above we
have introduced a cut off in the long range dipolar interaction. A
finite size scaling analysis of the magnetization, specific heat,
susceptibilities and Binder's cumulant clearly indicates that the
line labelled $a$ is of first order and the line $c$ is of second
order in agreement with other results. However, the $b$ line is of
$BKT$ type. After analysing the results obtained,  some questions
come out:
\begin{enumerate}
 \item Is it possible the existence of a limiting range of interaction
 in the dipolar term beyond which the character of the transition
 changes from $BKT$ to second order ?
 \item How does the line labelled $a$ end in the phase diagram ?
 \item What is the character of the intersection point of the
 three lines in the phase diagram ?
\end{enumerate}
In a very preliminary calculation Rapini et
al.\cite{rapini-costa-landau} studied the model with true dipolar
long range interactions. Their results led them to suspect of a
phase transition of the BKT involving the unbinding of
vortices-anti-vortices pairs in the model. However, to respond
those questions it is necessary to make an more detailed study of
the model for several values of the cut off range $r_c$. In a
simulation program we have to be careful in taking larger $r_c$
values since we have to augment the lattice size proportionally to
prevent misinterpretations.
\section{Acknowledgments}
Work partially supported by CNPq (Brazilian agency). Numerical
work was done in the LINUX parallel cluster at the Laborat\'{o}rio
de Simula\c{c}\~{a}o Departamento de F\'{i}sica - UFMG.
\end{document}